\documentclass[prd,twocolumn,amsmath,amssymb]{revtex4}
\usepackage{graphicx}

\def\al{\alpha}
\def\be{\beta}

\def\de{\delta}
\def\ep{\epsilon}

\def\ze{\zeta}
\def\et{\eta}
\def\th{\theta}
\def\vt{\vartheta}

\def\ka{\kappa}
\def\la{\lambda}

\def\rh{\rho}

\def\si{\sigma}
\def\vs{\varsigma}
\def\ta{\tau}

\def\ph{\phi}
\def\vp{\varphi}
\def\ch{\chi}
\def\ps{\psi}
\def\om{\omega}
\def\Ga{\Gamma}

\def\Ps{\Psi}
\def\Om{\Omega}
\def\pt#1{\phantom{#1}}
\def\prt{\partial}
\def\cl{{\cal L}}
\def\half{\tfrac12}
\def\qrt{\tfrac14}

\newcommand{\beq}{\begin{equation}}
\newcommand{\eeq}{\end{equation}}
\newcommand{\bea}{\begin{eqnarray}}
\newcommand{\eea}{\end{eqnarray}}
\newcommand{\rf}[1]{(\ref{#1})}
\def\etal{{\it et al.}}

\def\mn{{\mu\nu}}

\def\rs{{\rh\si}}

\def\kl{{\ka\la}}

\DeclareMathOperator{\re}{Re}
\DeclareMathOperator{\im}{Im}

\DeclareMathOperator{\sgn}{sgn}

\def\syjm#1#2{{}_{#1}Y_{#2}}

\def\q{q}
\def\qHat{\widehat{\q}}
\def\qd#1{{\q}^{(#1)}}

\def\s{s}
\def\sHat{\widehat{\s}}
\def\sd#1{{\s}^{(#1)}}

\def\k{k}
\def\kHat{\widehat{\k}}
\def\kd#1{{\k}^{(#1)}}

\def\kjm#1#2#3{k^{(#1)}_{(#2)#3}}
\def\kI{\kjm{d}{I}{jm}}
\def\kE{\kjm{d}{E}{jm}}
\def\kB{\kjm{d}{B}{jm}}
\def\kV{\kjm{d}{V}{jm}}
\def\kIdjm#1#2{\kjm{#1}{I}{#2}}
\def\kEdjm#1#2{\kjm{#1}{E}{#2}}
\def\kBdjm#1#2{\kjm{#1}{B}{#2}}
\def\kVdjm#1#2{\kjm{#1}{V}{#2}}

\def\kN{k^{{\rm N}(d)}_{jm}}

\def\cIdjm#1#2{c^{(#1)}_{(I)#2}}

\def\hb{\bar h}
\def\hs{\stackrel{\leftrightarrow}{h}}
\def\h#1{h_{(#1)}}
\def\F#1{F_{(#1)}}
\def\Y#1{{\mathcal Y}_{(#1)}}
\def\D{{\mathcal F}}

\def\ut{{\mathcal U}}
\def\At{{\mathcal A}}
\def\pt{{\Ps}}

\begin{document} 

\title{Signals for Lorentz violation in gravitational waves}
\author{Matthew Mewes}
\address{Physics Department, California Polytechnic State University, 
  San Luis Obispo, California 93407, USA}

\begin{abstract}
  Lorentz violations
  in gravitational waves are investigated.
  Plane-wave solutions for
  arbitrary gauge-invariant violations
  in linearized gravity are constructed.
  Signatures of Lorentz violation include
  dispersion, birefringence, and anisotropies.
  Modifications to waves from coalescing compact binaries
  and to strain signals in gravitational-wave detectors
  are derived.  
\end{abstract}

\maketitle

\section{Introduction}

The growing catalog of gravitational-wave observations
\cite{ligo1,ligo2,ligo3,ligo4,ligo5,ligo6}
offers new opportunities for tests of fundamental physics.
Lorentz invariance is a feature of both Einstein's general relativity
and the standard model of particle physics,
so any breaking of this symmetry
would signal new physics
\cite{reviews,tables,sme,smegrav}
potentially rooted in quantum gravity
\cite{strings}.
Because of the large propagation distances,
gravitational waves produced in binary mergers
are particularly sensitive to defects in relativity,
enabling new precision tests of Lorentz invariance in gravity
\cite{ligo3,kmgw1,ligodisp,speed1,speed2,disp0,disp1,disp2,%
  wang,cerenkov1,cerenkov2}.

Lorentz invariance is the combination
of both rotation symmetry and boost symmetry,
so Lorentz violations generally
produce unexpected directional
and velocity dependences.
One consequence is modified kinematics
for particles and waves.
However, 
to fully characterize
the effects of Lorentz violation
one needs a complete
dynamical model of the system.
A theoretical framework known as
the Standard-Model Extension (SME)
characterizes general violations
of Lorentz and CPT invariance
in both general relativity
and the standard model
at attainable energies
\cite{sme,smegrav}.
A Lorentz-violating term in the SME action
is formed from the contraction of
a conventional tensor operator with
a tensor coefficient for Lorentz violation.
The terms are classified according
to the mass dimension $d$
of the operator in natural units with $\hbar=c=1$
\cite{km09,nonmin}.
It is generally assumed that higher-$d$ terms
represent higher-order corrections
to conventional physics.

While particle sectors of the SME
have received intense scrutiny
over the last two decades \cite{tables},
fewer searches for Lorentz violation
in gravity have been performed.
Tests of Lorentz violation in gravitational waves
include searches for
birefringence \cite{kmgw1}
and dispersion \cite{disp2}.
Other tests of Lorentz invariance
in the SME gravity sector
include those involving
gravitational \v Cerenkov radiation \cite{cerenkov2},
atomic interferometers \cite{atomint},
superconducting gravimeters \cite{gravimeter},
orbital dynamics \cite{gpb,lunar,planetary,pulsars},
short-range-gravity experiments \cite{shortrange1,shortrange2},
comagnetometers \cite{comag},
nuclear binding energy \cite{benergy},
and very-long-baseline interferometry \cite{vlbi}.
The SME also serves as the foundation
for a number of theoretical studies
of Lorentz violation in gravity
\cite{smemattergrav,smepngrav,theory1,theory2,kmgw2,kmnewt}.

The development of the gravity sector of the SME
has progressed along several parallel lines,
each corresponding
to a different limit of the theory.
The most general extension
describes violations in gravity and particles,
including gravitational couplings
to standard-model fields \cite{smegrav}.
It is based on Riemann-Cartan geometry,
where the vierbein $e_\mu{}^a$ is the gravitational field.
One can then focus on matter-gravity couplings
\cite{smemattergrav}
or on the pure-gravity sector
\cite{smepngrav}.
The pure-gravity limit
assumes Riemannian geometry.
The usual Einstein-Hilbert action
is then augmented by all possible
coordinate-independent terms
involving the metric $g_\mn$.

The above construction produces an effective field theory
that encompasses all realistic violations
of Lorentz invariance in gravity,
whether they are explicit or dynamically generated
\cite{theory1}.
However,
the difficulties of working in a nonlinear theory
like general relativity are only exacerbated
by the inclusion by Lorentz violation,
encumbering systematic studies.
We can avoid many of the complications
by working at the level of linearized gravity \cite{lingrav}.
In this limit of the SME,
one posits that gravity is suitably weak
and expands the metric $g_\mn = \et_\mn+h_\mn$
about the constant Minkowski metric $\et_\mn$.
The action is constructed from all possible
Lorentz-invariant and Lorentz-violating terms
quadratic in the metric perturbation $h_\mn$.

The full linearized-gravity extension
is constructed in Refs.\ \cite{kmgw1,kmgw2}.
It takes the form of
an effective field theory in flat spacetime,
making its development and application
comparatively simple.
The action contains the linearized limit of
all Lorentz and diffeomorphism violations
in general relativity,
including those that violate the usual
gauge invariance,
$h_\mn \rightarrow h_\mn + \prt_{(\mu} \xi_{\nu)}$.
Work involving the linear extension
includes studies of gravitational-wave
dispersion relations for gauge-invariant
\cite{kmgw1}
and gauge-breaking violations
\cite{kmgw2}
and studies of Lorentz violation
in newtonian gravity
\cite{kmnewt}.

Some of the tightest constraints on
Lorentz violation in any sector
come from observations of
radiation from astrophysical sources,
where tiny modifications to the dynamics
can accumulate over cosmological times.
For example,
searches for photon dispersion
\cite{photondisp,kmapjl,friedman},
photon birefringence
\cite{kmapjl,friedman,photonbire},
and unconventional \v Cerenkov radiation
\cite{cerenkov3,cerenkov4}
have all placed tight limits on
particle-sector Lorentz violation.

In gravitational waves,
dispersion causes a deformation
of the waveform,
and birefringence causes changes
in the polarization.
Both will distort the strain signal
measured by gravitational-wave observatories.
This paper characterizes
dispersion and birefringence
due to Lorentz violation
and the effects of Lorentz violation
on gravitational waves
produced in the coalescence of compact binaries.

We restrict attention to the gauge-invariant
linearized-gravity sector of the SME \cite{kmgw1}.
We find that this limit produces
two independent modes
for propagation with differing phase velocities
and conventional polarizations
at zeroth order in Lorentz violation.
Note, however, that the full SME
includes gauge-breaking terms,
which are expected to produce effects
beyond those discussed here.
For example,
new exotic modes may propagate
that could be detected in
observations of gravitational waves
\cite{ligo5,extramodes}.

Throughout this work,
we adopt units with $c=1$,
but explicitly include Newton's constant $G_N$.
Setting  $G_N=1$ yields geometrized units.
Alternatively,
setting $\hbar=1$ gives natural units
and $G_N = 1/M_\text{Pl}^2$,
where $M_\text{Pl}$ is the Planck mass.
Spacetime indices are raised and lowered
using the Minkowski metric with
$(-,+,+,+)$ signature.

This paper is organized as follows.
Section \ref{sec_gw}
examines gravitational plane waves
in gauge-invariant linearized gravity
with Lorentz violation.
In Sec.\ \ref{sec_bin},
we consider waves created in binary mergers
and derive the modified detector strain 
including the effects of Lorentz violation
during propagation.
Section \ref{sec_disc} provides
a summary and some concluding remarks.
The effects of Lorentz violation on
gravitational Stokes parameters
are discussed in the Appendix.

\section{GRAVITATIONAL WAVES}
\label{sec_gw}

In this section,
we find plane-wave solutions for gravitational waves
in the presence of Lorentz violation.
We begin by first reviewing
the gauge-invariant linearized-gravity
sector of the SME.
We then derive the leading-order dispersion relation
and the polarizations of the propagating modes.
The effects on waves that have traveled
astrophysical distances are explored,
and several special cases are discussed.

\subsection{Basic theory}
\label{sec_gw_theory}

The lagrangian for the gauge-invariant
linearized-gravity sector of the SME
consists of all possible terms quadratic in $h_\mn$,
including arbitrary numbers of derivatives of $h_\mn$.
It contains the usual linearized Einstein-Hilbert lagrangian
and an infinite series of Lorentz-invariant
and Lorentz-violating terms.
It can be written in the compact form \cite{kmgw1}
\bea
\cl &=&
\tfrac14 \ep^{\mu\rh\al\ka} \ep^{\nu\si\be\la} 
\et_\kl h_\mn \prt_\al\prt_\be h_\rs
\notag\\
&&+ \tfrac14 h_\mn 
(\sHat{}^{\mu\rh\nu\si} 
+ \qHat{}^{\mu\rh\nu\si} 
+ \kHat{}^{\mu\nu\rh\si})
h_\rs \ .
\label{lagrangian}
\eea
Each term is invariant under
the usual gauge transformation
$h_\mn \rightarrow h_\mn + \prt_{(\mu} \xi_{\nu)}$
up to a total derivative.
The first line in Eq.\ \rf{lagrangian}
is the conventional lagrangian and generates
the usual linearized Einstein tensor
$G^\mn = -\half \et_\rs\ep^{\mu\rh\al\ka}\ep^{\nu\si\be\la}
\prt_\al \prt_\be h_\kl$.

The last term in Eq.\ \rf{lagrangian}
contains the Lorentz-violating modifications.
It naturally splits into three different classes of violations,
corresponding to the operators
\bea
\sHat{}^{\mu\rh\nu\si} &=&
\sum \sd{d}{}^{\mu\rh\al_1\nu\si\al_2\ldots\al_{d-2}}
\prt_{\al_1}\ldots \prt_{\al_{d-2}} \ ,
\notag \\
\qHat{}^{\mu\rh\nu\si} &=&
\sum \qd{d}{}^{\mu\rh\al_1\nu\al_2\si\al_3\ldots\al_{d-2}}
\prt_{\al_1}\ldots \prt_{\al_{d-2}} \ ,
\notag \\
\kHat{}^{\mu\nu\rh\si} &=&
\sum \kd{d}{}^{\mu\al_1\nu\al_2\rh\al_3\si\al_4\ldots\al_{d-2}}
\prt_{\al_1}\ldots \prt_{\al_{d-2}}\ .
\label{sqk}
\eea
The tensor coefficients in these expansions
control the Lorentz violation.
Each has different symmetries,
which are summarized in Table 1 of Ref.\ \cite{kmgw1}.
The $\s$- and $\k$-type violations are CPT even,
while $\q$-type violations break CPT invariance
in addition to Lorentz invariance.
For particles, CPT breaking typically leads
to different properties for particles
and antiparticles.
For gravitational waves,
CPT violation breaks the degeneracy
between left- and right-handed polarizations.
The sums in Eq.\ \rf{sqk} are over
even $d\geq4$ for $\s$-type violations,
odd $d\geq5$ for $\q$-type, and
even $d\geq6$ for $\k$-type.

The equations of motion for
Eq.\ \rf{lagrangian} can be written in the form
\beq
0=G^{\mn} + \de M^{\mn\rs}h_\rs \ ,
\label{eom}
\eeq
where the tensor operator
\bea
\de M^{\mn\rs}
&=& -\qrt\big(\sHat{}^{\mu\rh\nu\si}
+\sHat{}^{\mu\si\nu\rh}\big)
-\half \kHat{}^{\mu\nu\rh\si}
\notag \\ &&
-\tfrac{1}{8}\big(
 \qHat{}^{\mu\rh\nu\si}
+\qHat{}^{\nu\rh\mu\si}
+\qHat{}^{\mu\si\nu\rh}
+\qHat{}^{\nu\si\mu\rh}\big) \
\label{deM}
\eea
contains the unconventional parts.
This operator is symmetric in the first pair
of indices and the last pair of indices.
The CPT-even part of $\de M^{\mn\rs}$
is symmetric under interchange of the first
and last pair of indices
and involves an even number of derivatives.
The CPT-odd part is antisymmetric
under interchange of the pairs of indices
and contains an odd number of derivatives.
Consequently,
$\de M^{\mn\rs}$ is a hermitian operator
acting on the space of symmetric rank-2 tensors.

\subsection{Eigenmodes}
\label{sec_gw_modes}

We next derive leading-order
plane-wave solutions of
the equations of motion.
A fourier transform converts
$\prt_\al \rightarrow ip_\al$
and Eq.\ \rf{eom} to
a $p$-dependent matrix equation.
The operators 
$\sHat{}^{\mu\rh\nu\si}$,
$\qHat{}^{\mu\rh\nu\si}$,
and $\kHat{}^{\mu\nu\rh\si}$
are now interpreted as functions of $p_\al$.
Solving the $p$-space equations of motion gives
plane-wave solutions
with wave vector $p^\al = (\om;\vec p\,)$.
To handle both positive and negative frequencies,
it is useful to write $\vec p = \om\hat v/v$.
The unit vector $\hat v = \sgn(\om) \vec p/|\vec p|$
points in the direction of propagation,
and $v=|\om|/|\vec p|$ is the phase velocity.
Note that $\hat v$
points in the direction of $\vec p$ for positive frequencies
and opposite to $\vec p$ for negative frequencies.

In the usual case,
one normally starts in a Hilbert gauge
by imposing the Lorenz condition $p_\nu\hb^{\mu\nu} = 0$.
The Einstein equation then reads
$G^\mn = \half p^2 \hb^\mn = 0$,
giving the dispersion relation $p^2=0$.
On shell, you can choose a gauge that
is temporal ($h^{0\mu}=0$),
transverse ($h^{jk} p^k = 0$),
and traceless ($h^{jj}=0$),
leading to the transverse traceless gauge.
Two degrees of freedom remain,
giving two degenerate polarizations.

In the Lorentz-violating case,
where the on-shell $p_\al$ is no longer light-like,
we cannot necessarily impose all
of the above gauge conditions simultaneously.
It is convenient, however,
to work in the temporal gauge.
The other gauge conditions may or may not
be satisfied on shell.
An advantage of the temporal gauge
is that coordinates for free nonrelativistic test masses
are inertial, since
$\prt_t^2 x^j \approx -\Ga^j_{00} = 0$.

It is also useful to work
in a helicity basis with basis vectors \cite{km09}
\bea
\hat e_r &=& \hat e^r
= \hat v = (\sin\th\cos\ph,\sin\th\sin\ph,\cos\th) ,
\notag \\
\hat e_\pm &=& \hat e^\mp
= \tfrac{1}{\sqrt{2}}(\hat e_\th \pm i\hat e_\ph) \ .
\eea
The propagation vector
$\hat v$ defines the ``radial'' direction,
and $\hat e_\th$ and $\hat e_\ph$
are the usual unit vectors associated with
spherical-coordinate angles $\th$ and $\ph$.
The complex helicity vectors $\hat e_\pm$ span
the transverse subspace.

In the temporal gauge,
only the spatial parts of $h_\mn$ are nonzero,
which we write as
$\hs = h^{ab} \hat e_a\otimes\hat e_b$
in terms of helicity-basis components
$h^{ab} = \hat e^a\cdot\!\!\hs\!\!\cdot\,\hat e^b$.
Note that raising and lowering spatial indices
in the helicity basis is done using
the skew-diagonal helicity metric
$\et_{ab} = \et^{ab} = \hat e_a\cdot\hat e_b$.
The result is that raising or lowering
helicity-basis indices changes $\pm$ to $\mp$.
For example, $G^{r+}=G_{r-}$.

The ten helicity components of
the $p$-space Einstein tensor can be written
in terms of the six components of
the temporal-gauge $\hs$.
The unconventional $0$-helicity components
are given by
$G^{rr} = v G^{0r} = v^2 G^{00} = \om^2 h^{+-}$
and
$G^{+-} = \tfrac{\om^2}{2}h^{rr} - \tfrac{p^2}{2} h^{+-}$.
The $\pm1$-helicity components obey
$G^{r\pm} = v G^{0\pm} = -\tfrac{\om^2}{2} h^{r\pm}$.
The transverse $\pm2$-helicity components give
$G^{\pm\pm} = \tfrac{p^2}{2} h^{\pm\pm}$.
In the usual case,
where $G^\mn=0$ and $p^2=0$,
only the $h^{\pm\pm}$ components can be nonzero,
giving an $\hs$ that is transverse and traceless.
In the Lorentz-violating case,
we can use the above relations
to construct perturbative solutions.

Assuming the $0$-helicity and $\pm1$-helicity
components of $\hs$ are small,
the leading-order $\pm2$-helicity
components $h^{\pm\pm}$
satisfy the matrix equation
\bea
\begin{pmatrix}
p^2 + 2\de M^{++--} & 2\de M^{++++}\\
2\de M^{----} & p^2 +2\de M^{--++}
\end{pmatrix}
\begin{pmatrix}
  \h{+2}\\\h{-2}
\end{pmatrix}
= 0 \ ,
\label{eom2}
\eea
where we denote
\beq
\h{\pm2} = h^{\pm\pm} \ .
\eeq
After solving for the leading-order $\h{\pm2}$,
we can use them and
the modified Einstein equation \rf{eom}
to perturbatively solve for
higher-order corrections to the polarization.
This procedure leads to a temporal-gauge $h^\mn$
that differs from a conventional $h^\mn$
by corrections that are suppressed by
coefficients for Lorentz violation.
Note that the result is neither
transverse nor traceless.
However, the unconventional parts
are likely to be too small to be directly observable.
Therefore, the dominant effect of Lorentz violation
on the polarization is
a possible breaking of the usual degeneracy
between the two polarizations,
resulting in birefringence.

Each of the matrix elements in Eq.\ \rf{eom2}
has definite helicity.
The diagonal elements of
the square matrix have zero helicity,
while the off-diagonal elements have helicity $\pm4$,
coupling the right-handed $\h{+2}$
and left-handed $\h{-2}$ polarizations.
The trace element preserves
the usual degeneracy between
the two polarizations,
motivating the following definition
\bea
\vs^0 &=& -\tfrac{1}{2\om^2}\big(\de M^{++--}+\de M^{--++}\big)
\notag \\
&=& \tfrac{1}{2\om^2}\big(\sHat^{+-+-}+\kHat^{++--}\big) \ .
\eea
We expect the remaining parts to
break the degeneracy,
giving birefringence.
We define these as
\bea
\vs_{(0)} &=& \tfrac{1}{2\om^2}\big(\de M^{++--}-\de M^{--++}\big)
= -\tfrac{1}{2\om^2} \qHat^{+-+-} \ ,
\notag \\
\vs_{(\pm 4)} &=& \tfrac{1}{2\om^2}\de M^{\pm\pm\pm\pm}
= -\tfrac{1}{2\om^2} \kHat^{\pm\pm\pm\pm} \ .
\eea
The combinations $\vs^0$ and $\vs_{(0)}$
are real and have zero helicity,
while $\vs_{(\pm4)}$ have helicity $\pm4$
and obey $\vs_{(\pm4)}^* = \vs_{(\mp 4)}$.

In this work
the $\vs$ functions are found
by fixing the gauge
and working in the helicity basis.
Note, however,
that covariant versions of these functions
and the dispersion relation
are derived in Ref.\ \cite{kmgw1}
without gauge fixing using
the methods discussed in
Ref.\ \cite{kmgw2}.

As discussed in the Appendix,
the $p$-dependent coefficient combinations
$\vs^0$, $\vs_{(0)}$, $\vs_{(+4)}$, $\vs_{(-4)}$
can be interpreted as conveniently normalized
Stokes parameters for the faster propagating mode.
They are functions of the frequency $\om$
and the wave vector $\vec p$.
However, when evaluating the $\vs$ functions,
we can assume the usual energy-momentum relation
and take $\vec p = \om\hat v$
at leading order.
We then get functions that
depend on the frequency $\om$
and propagation direction $\hat v$.
They can be written as
\bea
\vs^0
&=& \sum_d \om^{d-4} \vs^{(d)0}(\hat v) \ ,\notag \\
\vs_{(\pm4)} 
&=& \sum_d \om^{d-4} \vs^{(d)}_{(\pm4)}(\hat v) \ , \notag \\
\vs_{(0)}
&=& \sum_d \om^{d-4} \vs^{(d)}_{(0)}(\hat v) \ ,
\label{sigma1}
\eea
separating the frequency and direction
dependences.

The direction-dependent factors can be
expanded in spin-weighted spherical harmonics
$\syjm{s}{jm}$.
Spin weight is the opposite of helicity \cite{km09},
so the expansions take the form \cite{kmgw1}
\bea
\vs^{(d)0}(\hat v) &=&
\sum_{jm} (-1)^j\, \syjm{0}{jm}(\hat v)\, \kI \ , \notag \\
\vs^{(d)}_{(\pm4)}(\hat v) &=&
\sum_{jm}(-1)^j\, \syjm{\mp4}{jm}(\hat v)\, (\kE \pm i \kB) \ , \notag \\
\vs^{(d)}_{(0)}(\hat v) &=&
\sum_{jm}(-1)^j\, \syjm{0}{jm}(\hat v)\, \kV \ .
\label{sigma2}
\eea
The spherical coefficients for Lorentz violation 
$\kI$, $\kV$, $\kE$ and $\kB$
are linear combinations of the underlying
tensor coefficients in Eq.\ \rf{sqk}.
The connection places limits on
the angular momentum indices $j$ and $m$
for each dimension $d$.
These limits, along with the coefficient count,
are given in Table \ref{kcoeffs}.
The spherical coefficients for Lorentz violation
have mass dimension $4-d$
in units with $\hbar=1$.
In geometrized units,
they have length dimension $d-4$.
Note that $(-1)^j\, \syjm{s}{jm}(\hat v) = \syjm{-s}{jm}(-\hat v)$,
which is convenient in astrophysical tests
where $-\hat v$ gives the location of the source.

\begin{table}
  \setlength{\tabcolsep}{7pt}
  \begin{tabular}{cccc}
    Coefficient & $d$ & $j$ & Number\\
    \hline
    \hline
    $\kI$ & $\text{even},\geq4$ & $0,1,\ldots,d-2$ &$(d-1)^2$\\
    $\kV$ & $\text{odd},\geq5$  & $0,1,\ldots,d-2$ &$(d-1)^2$\\
    $\kE$ & $\text{even},\geq6$ & $4,5,\ldots,d-2$ &$(d-1)^2-16$\\
    $\kB$ & $\text{even},\geq6$ & $4,5,\ldots,d-2$ &$(d-1)^2-16$\\
  \end{tabular}
  \caption{\label{kcoeffs}
    Summary of the spherical coefficients for Lorentz violation \cite{kmgw1}.
    The second and third columns give the ranges for the dimension
    index $d$ and angular-momentum index $j$.
    The $m$ index obeys the usual relation $-j\leq m\leq j$.
    The last column gives the total number of independent coefficients for each $d$.
    Each set of coefficients obeys the complex-conjugation relation
    $k_{jm}^{(d)*}=(-1)^m k_{j(-m)}^{(d)}$.
  }
\end{table}

Different physical systems access different
linear combinations of
the fundamental coefficients \cite{kmnewt}.
The spherical coefficients in Eq.\ \rf{sigma2}
represent the subset affecting gravitational waves
at leading order.
Note, however, that a given point source
with fixed observed $\hat v$
can at most measure the four
linear combinations of spherical coefficients
given in Eq.\ \rf{sigma2}.
Different sources with different $\hat v$
will access different linear combinations.
One can in principle disentangle
the numerous spherical coefficients
for Lorentz violation at any dimension $d$
by combining data from multiple sources
at different locations on the sky.

Nontrivial solutions to Eq.\ \rf{eom2}
exist when the determinant of the $2\times2$
matrix vanishes.
This gives the dispersion relation
\beq
p^2 = 2\om^2\big(\vs^0 \mp |\vec\vs|\big) \ ,
\label{dr}
\eeq
where we define
\beq
|\vec\vs\,| = \sqrt{\big|\vs_{(+4)}\big|^2 + \big|\vs_{(0)}\big|^2} \ .
\eeq
Solving for the frequency,
the dispersion relation can be written as
$|\om| = \big(1-\vs^0 \pm |\vec\vs\,|\big)|\vec p|$,
giving phase velocities
\beq
v_{\pm} = 1-\vs^0
\pm |\vec\vs\,| \ .
\label{vph}
\eeq
The usual degeneracy between
the polarizations is broken when $|\vec\vs\,|\neq 0$,
as expected,
and the two modes propagate at different speeds.
The top sign in these expressions corresponds
to the fast mode,
and the bottom sign gives the slow mode.

To find the polarization of each mode,
we solve Eq.\ \rf{eom2} on shell.
The result can be written in terms of two angles
$\vt$ and $\vp$ that completely characterize
the polarizations of the modes.
They are defined through
\beq
\sin\vt = \frac{\big|\vs_{(+4)}\big|}{|\vec\vs\,|} \ , \quad
\cos\vt = \frac{\vs_{(0)}}{|\vec\vs\,|}\ , \quad
e^{\mp i\vp}  =
\frac{\vs_{(\pm 4)}}{\big|\vs_{(+4)}\big|} \ .
\label{bire_angles}
\eeq
We then find that
the fast mode has normalized polarization
\beq
\begin{pmatrix}
\h{+2}\\ \h{-2}
\end{pmatrix}_\text{\!\!fast}
=
\begin{pmatrix}
\cos\tfrac\vt2 e^{-i\vp/2}\\
\sin\tfrac\vt2 e^{i\vp/2} 
\end{pmatrix} \ ,
\label{fast}
\eeq
while the polarization of the slow mode
can be written
\beq
\begin{pmatrix}
  \h{+2}\\ \h{-2}
\end{pmatrix}_\text{\!\!slow}
=
\begin{pmatrix}
-\sin\tfrac\vt2 e^{-i\vp/2}\\
\cos\tfrac\vt2 e^{i\vp/2} 
\end{pmatrix} \ .
\label{slow}
\eeq
A general polarization is a linear
combination of the two eigenmodes.
The unitary transformation
\bea
\begin{pmatrix}
\h{+2}\\ \h{-2}
\end{pmatrix}
=
\begin{pmatrix}
\cos\tfrac\vt2 e^{-i\vp/2} &&
-\sin\tfrac\vt2 e^{-i\vp/2} \\
\sin\tfrac\vt2 e^{i\vp/2} &&
\cos\tfrac\vt2 e^{i\vp/2}
\end{pmatrix}
\begin{pmatrix}
\h{f}\\ \h{s}
\end{pmatrix}\quad
\label{eigenmodes}
\eea
relates the helicity components
$\h{\pm2}$ of an arbitrary polarization
to its fast-mode component $\h{f}$
and its slow-mode component $\h{s}$.

Combined with the dispersion relation \rf{dr},
the above polarizations
describe the leading-order
effects in gravitational waves
for any gauge-invariant extension
to linearized gravity,
including all possible Lorentz-violating
and Lorentz-invariant modifications.

\subsection{Dispersion and birefringence}
\label{sec_gw_lv}

The unconventional parts of the phase velocity
lead to a gradual shift in phase
as the wave propagates.
Consider, for example,
a simple plane wave in flat spacetime
that has propagated a distance $l$.
For a deformed phase velocity $v=1+\de v$,
we get 
$h(t) \sim e^{-i\om(t-l/v)}
\approx e^{-i\om \de v l} e^{-i\om(t-l)}$,
shifting the phase by $\om \de v l$.

For cosmological sources in an expanding universe,
the redshift of the frequency can be accounted for
by considering an infinitesimal change in the phase,
$d\psi = d\psi_0 + \om\de v dl$.
Integrating from the source to the observer,
the first part gives the conventional phase.
The second part gives the Lorentz-violating contribution,
$\de\psi_\pm = \int dl\, \om (-\vs^0 \pm |\vec\vs\,|)$,
for the fast and slow modes.
At zeroth order, the wave propagates at $v=1$,
so we can replace the distance interval $dl$
with the propagation time $dt = -dz/(1+z)/H(z)$,
where $H(z)$ is the Hubble expansion rate at redshift $z$.
The accumulated Lorentz-violating phase is then given by
\beq
\de\ps_\pm =
\om \int_0^z dz\,
\frac{-\vs^0 \pm |\vec\vs\,|}{H(z)}
= -\de \pm \be \ ,
\label{phase}
\eeq
where $\om$ is the observed frequency.
The common phase $\de$ is independent of polarization
and leads to dispersion but no birefringence.
The birefringent phase $\be$ is the polarization dependent,
causing the net polarization to evolve as the wave propagates.
For fixed dimension $d$,
we can write the phases as
\beq
\de = \om^{d-3}\, \ta\, \vs^{(d)0} \ , \quad
\be = \om^{d-3}\, \ta\, \big|\vec\vs\,{}^{(d)}\big| \ ,
\label{lvphases}
\eeq
where
\beq
\ta = \int_0^z  \frac{(1+z)^{d-4}}{H(z)}\, dz
\eeq
is an effective $d$-dependent propagation time
that accounts for the redshift in $\om$ during propagation.

As the wave propagates,
the phase of the fast and slow components
shifts relative to the conventional case,
leading to observed components
$\h{f,s} = e^{i\de \mp i\be} \h{f,s}^\text{LI}$,
where $\h{f,s}^\text{LI}$ is the Lorentz invariant limit.
Using Eq.\ \rf{eigenmodes},
we transform this result to the helicity basis,
giving
\bea
\h{\pm2} &=&
e^{i\de}
(\cos\be \mp i\cos\vt\sin\be)
\h{\pm2}^\text{LI}
\notag \\ &&
-ie^{i\de}\sin\vt e^{\mp i\vp} \sin\be
\h{\mp2}^\text{LI} \ .
\label{htrans}
\eea
We can also write this
in terms of standard ``plus'' and ``cross''
linear polarizations,
defined as
$\h+ = h^{\th\th} = -h^{\ph\ph}$ and
$\h\times = h^{\th\ph} = h^{\ph\th}$.
These are related to the helicity components through
\beq
\h{\pm2} = \h+ \mp i \h\times \ .
\label{pols}
\eeq
The changes to the linear polarizations are
given by
\bea
\h+ &=& e^{i\de}(\cos\be-i\sin\vt\cos\vp\sin\be)\h{+}^\text{LI}
\notag \\ &&\qquad
- e^{i\de}(\cos\vt+i\sin\vt\sin\vp) \sin\be\h\times^\text{LI} \ ,
\notag\\
\h\times &=& e^{i\de}(\cos\be+i\sin\vt\cos\vp\sin\be)\h\times^\text{LI}
\notag \\ &&\qquad
+ e^{i\de}(\cos\vt-i\sin\vt\sin\vp) \sin\be\h+^\text{LI} \ .\qquad
\label{htranslin}
\eea
Equations \rf{htrans} and \rf{htranslin}
give the predicted effects for
general modifications to linearized gravity.
They provide a map between the modified theory
and the conventional limit
and incorporate dispersive changes in phase
and changes in polarization due to birefringence.

While the above applies
to general cases,
one simplifying strategy is
to consider various special limits.
The three main classes of Lorentz violation
in gravitational waves are nonbirefringent violations,
CPT-odd birefringent violations,
and CPT-even birefringent violations.
We briefly consider each of these in turn.

{\em Nonbirefringent violations.}
The CPT-even $\kI$ coefficients are responsible for
nonbirefringent Lorentz violations
and exist for even $d\geq4$.
They generally produce a frequency-dependent
phase velocity producing dispersion.
Note, however, that the $d=4$
case gives a phase velocity that depends
on direction but is frequency independent.
Consequently, only $d\geq6$ violations produce dispersion.
In this limit, the birefringent phase $\be$ vanishes,
and Eqs.\ \rf{htrans} and \rf{htranslin} reduce to
$\h{\cdot} = e^{i\de} \h{\cdot}^\text{LI}$
for all polarizations,
giving a change in phase
but no change in polarization.

{\em CPT-odd birefringence.}
The $\kV$ coefficients give both dispersion
and birefringence and exist for odd $d\geq5$.
Setting all other coefficients to zero,
we find that the eigenmodes are circularly polarized.
We get $\vt=0$ when
the right-handed polarization $\h{+2}$ is faster
and $\vt=\pi$ when
the left-handed polarization $\h{-2}$ is faster.
In both cases,
the circular polarizations acquire a simple phase shift,
$\h{\pm2} = \h{\pm2}^\text{LI} e^{\mp i \de\ps}$,
where $\de\ps = \om^{d-3}\ta\vs_{(0)}^{(d)}$.
The shift in relative phase between
the two circular polarizations causes
a rotation of the linear polarizations:
\beq
\begin{pmatrix}
  \h+ \\ \h\times
\end{pmatrix}
=
\begin{pmatrix}
  \cos\de\ps & -\sin\de\ps\\
  \sin\de\ps & \cos\de\ps
\end{pmatrix}
\begin{pmatrix}
  \h{+}^\text{LI} \\ \h{\times}^\text{LI}
\end{pmatrix} \ .
\eeq
This corresponds to a simple rotation of $\hs$
about $\hat v$ by angle $\de\ps/2$,
leaving the degree of linear
and circular polarization unchanged.
Note that the polarization will remain
fixed if the wave is produced in one of
the circularly polarized eigenmodes,
but frequency dependence in
the phase velocity still produces dispersion.

{\em CPT-even birefringence.}
The $\kE$ and $\kB$ coefficients also
give dispersion and birefringence,
but exist for even $d\geq6$.
The changes in the wave
can be found by setting $\de=0$ and $\vt=\pi/2$
in Eqs.\ \rf{htrans} and \rf{htranslin}.
The result is more complicated in this case
because the eigenmodes are linearly polarized
at polarization angles
$\vp/4$ and $\vp/4+\pi/4$.
Only linearly polarized waves
at one of these angles
will maintain constant polarization.
All waves will experience dispersion due
to the frequency-dependent phase velocities.

The effects of birefringence
in the CPT-even case and more general cases
can be made more transparent
by considering the gravitational Stokes parameters,
as discussed in the Appendix.
In general, birefringence produces
a simple rotation of the Stokes vector
for the wave about the Stokes vector
for the faster eigenmode,
which can have any elliptical polarization.

Dimension $d=4$ violations do not affect chirp observations
since they are nondispersive and nonbirefringent.
In this case,
the phase and group velocities acquire the same
frequency- and polarization-independent shift
\beq
\de v = \sum (-1)^{j+1} \syjm{0}{jm}(\hat v)
\kIdjm{4}{jm} \ .
\eeq
While this doesn't affect chirp signals,
it can be tested through
time-of-flight comparisons with photons
\cite{speed1,speed2}.
For example, assuming a common origin
for GW170817 and GRB 170817A,
Ref.\ \cite{speed2} places limits on the difference
between the speed of gravity and the speed of light.
The result can be translated to a constraint
on a combination of spherical SME coefficients:
\beq
-3\times 10^{-15}\leq
\sum_{jm} \syjm{0}{jk}(\hat n)
\big(\cIdjm{4}{jm}-\kIdjm{4}{jm}\big)
\leq 7\times 10^{-16} \ ,
\eeq
where $\cIdjm{4}{jm}$ are photon-sector
coefficients for Lorentz violation.
Here we assume negligible birefringence in photons \cite{photonbire}.
Using the location of the optical counterpart \cite{GWGRB},
the source location $\hat n=-\hat v$
has angles
$\{\th_{\hat n},\ph_{\hat n}\}
\simeq \{113^\circ,197^\circ\}$
in the Sun-centered frame
used in tests involving the SME.
Restricting attention to the isotropic limit,
the shift in the speed of gravity reduces to 
$\de v = - \sqrt {1/4\pi} \kIdjm{4}{00}$.
Assuming isotropy in photons as well,
the constraint above gives
\beq
-11\times 10^{-15}
\lesssim
\cIdjm{4}{00}-\kIdjm{4}{00}
\lesssim
25\times 10^{-16} \ .
\eeq

\section{BINARY COALESCENCE}
\label{sec_bin}

The goal of this section is to apply
the above results to coalescing binaries
in order to characterize the signatures
of Lorentz violation in chirp signals.
We begin by reviewing
conventional mergers
in the quadrupole approximation,
which describes the dominate expected features
of the emitted gravitational waves
in the Lorentz-invariant case.
We then calculate the modifications
due to Lorentz violation.
General expressions for
Earth-incident gravitational waves
and detector strains are found
in terms of coefficients for Lorentz violation
in the standard Sun-centered celestial equatorial
reference frame used in Lorentz tests involving the SME.
We then discuss signatures
in several special limits.

\subsection{Conventional mergers}
\label{sec_bin_conv}

In order to search for Lorentz violation
in chirp signals produced by binary systems,
we first need a realistic
description of the expected wave
without Lorentz violation.
This section provides
a brief discussion of conventional binary mergers
and establishes the basic structure
needed to construct gravitational-wave
signals with Lorentz violation.
For a review of the underlying conventional physics see,
for example, Refs.\ \cite{will,pnbins}.

Starting in the time domain,
we imagine an asymptotically flat frame
centered on the merger.
We orient the frame so that
the binary revolves in the right-handed sense
about the $z$ axis.
Far from the merger,
the wave will be transverse
to the propagation direction $\hat v$
and the position vector $\vec x \approx r\hat v$.
The conventional gravitational radiation
is then completely described by either of
the complex helicity components,
since $\h{+2}(t) = \h{-2}^*(t)$
in the time domain,
where $h_\mn$ is real.

Conventional waves produced
by binaries are dominated by
their merger-frame $j=2,\ m=\pm2$ multipoles
and are even under parity.
This implies that the observed wave
takes the generic form
\beq
\h{+2}(t) = \ut(t)\, \syjm{-2}{22}(\hat v)
+ \ut^*(t)\, \syjm{-2}{2(-2)}(\hat v) \ .
\label{hyjmt}
\eeq
The wave is then completely characterized
by a single complex scalar function $\ut(t)$.

While the exact form of $\ut(t)$
can only be found by considering
the detailed physics of the merger,
some general features can be ascertained.
For example,
splitting $\ut(t)$ into an amplitude and a phase,
$\ut(t) = \At(t)e^{-i\pt(t)}$,
we expect both the phase $\pt(t)$
and its rate of change to increase monotonically with time.
To see this, we write $\h{+2}$
in terms of the spherical-coordinate
angles $\th$ and $\ph$
for the direction vector $\hat v$:
\beq
\h{+2}(t) = \sqrt{\tfrac{5}{4\pi}}\big(
\ut e^{2i\ph} \cos^4\tfrac{\th}{2}
+ \ut^* e^{-2i\ph} \sin^4\tfrac{\th}{2}\big) \ .
\eeq
We then note that $\ut$
only contributes
through the combination
$\ut e^{2i\ph}
= \At e^{i(2\ph-\pt)}$,
so wavefronts satisfy 
$2\ph = \pt + (\text{constant})$.
Since we assume right-handed
rotations about the $z$ axis,
we expect the azimuthal angle $\ph$ of a wavefront
at a fixed distance from the merger to increase with time,
so the phase function $\pt$
increases monotonically with time.
Also, since the rotation of
the binary accelerates as it inspirals,
we expect a positive instantaneous oscillation frequency
$\Om(t) = \prt_t{\pt}(t)$
that increases with time.
We also note that the wave frequency
$\Om$ is twice the wavefront rotation rate,
as expected in a binary system.

The frequency-domain version of
Eq.\ \rf{hyjmt} is more amenable
to studies of Lorentz violation.
Taking the fourier transform,
we get
\beq
\h{+2}(f) = u(f)\, \syjm{-2}{22}(\hat v)
+ u^*(-f)\, \syjm{-2}{2(-2)}(\hat v) \ ,
\label{hyjmf}
\eeq
where $u(f) = \int dt\, \ut(t) e^{i2\pi f t}$,
for $2\pi f = \om$.
The $u(f)$ function completely characterizes
the wave in the frequency domain.
The negative-helicity component can be found
using $\h{-2}(f) = \h{+2}^*(-f)$.
We also split the frequency-domain function
\beq
u(f) = A(f) e^{i\ps(f)}
\eeq
into an amplitude and a phase.

The connection between the time- and frequency-domains
can be studied through a stationary-phase approximation.
For a fixed frequency $f$,
the fourier transform
$u(f) = \int dt\, \At(t) e^{i(2\pi ft-\pt)}$
is dominated by times $t_f$
that are extrema of the phase.
These times are solutions to the equation
$2\pi f = \Om(t_f)$.
Since the instantaneous frequency $\Om(t)$ is positive,
extrema only exist for positive frequencies,
and we can assume that negative frequencies
play an insignificant role in $u(f)$.
The negative frequencies do affect
the signal through $u^*(-f)$
in Eq.\ \rf{hyjmf},
however.
Since $\Om$ increases with time,
we can in principle
invert the relationship between $f$ and $t_f$,
so $t_f$ is a single-valued function of $f$.
Then the usual stationary-phase approximation gives
$A(f)\approx \sqrt{2\pi/\prt_t^2{\pt}(t_f)}\,\At(t_f)$
and
$\ps(f) \approx 2\pi ft_f-\pt(t_f)-\tfrac\pi4$.

Applying the stationary-phase approximation
to the inverse fourier transform yields similar relations.
It implies that the dominant frequency $f_t$ at time $t$
is a solution of $2\pi t = \prt_f\ps(f_t)$.
Again, since $f_t$ and $t$ are expected to rise together,
we can assume $\prt_f^2\ps(f_t)$ is positive.
We then arrive at two key characteristics
of the frequency-domain function
$u(f) = A(f) e^{i\ps(f)}$.
To a good approximation,
we can assume that $u(f)$ is nonzero
for positive frequencies only
and that the phase function $\ps(f)$ is convex.

As a chirp progresses,
it transitions from
an inspiral to a merger,
followed by the ringdown.
Each stage is subject to different physics
and gives different contributions to
$A(f)$ and $\ps(f)$.
For example,
early in the inspiral
the ``newtonian'' approximation is valid
\cite{mtw},
which results in
\bea
\ps(f) &\approx&
2\pi f t_0 + \ps_0
+ \tfrac{3}{2^7\pi^{5/3}} \et^{-1}(Tf)^{-5/3} \ ,
\notag\\
A(f) &\approx&
C r^{-1} T^2\et^{1/2} (Tf)^{-7/6} \ ,
\label{newton}
\eea
where $C$ is a constant,
$r$ is the distance from the source,
$t_0$ determines the time origin,
and $\ps_0$ is a phase constant.
The important features of the chirp
depend on the total mass $M=M_1+M_2$
and mass ratio $\et=M_1M_2/M^2$.
Here we define the chirp time constant
$T = G_N (1+z)M$,
which gives the characteristic time scale.
We incorporate the source redshift $z$
to account for cosmological expansion.

The expressions in Eq.\ \rf{newton}
provide a simple approximation that
should hold at low frequencies.
At higher frequencies,
corresponding to later times,
the approximation is expected to fail.
More accurate descriptions at all frequencies
can be achieved through a combination of
higher-order post-newtonian corrections \cite{pnbins}
and numerical relativity \cite{nrbins}.
Analytic templates can be constructed to
approximate the late-stage physics.
These generally have a phase function of the form
\beq
\ps(f) = 2\pi ft_0 + \ps_0 + \sum_n (Tf)^{n/3}\ps_n \ ,
\eeq
where the $\ps_n$ are constants.
The amplitude $A(f)$ may also be modified.
For example, Ref.\ \cite{ajith1}
assumes an $A(f)$ that
is proportional to $f^{-7/6}$ over inspiral frequencies,
is proportional $f^{-2/3}$ for the merger,
and decays as a lorentzian at higher ringdown frequencies.
Their phase includes terms with $n = -5,-3,-2,-1,1,2$.
All parameters are treated as functions
of the mass ratio $\et$ and fit to results
from post-newtonian and numerical relativity.
The spins of the two bodies can also
be incorporated in the templates \cite{ajith2}.

\subsection{Signatures of Lorentz violation}
\label{sec_bin_lv}

The choice of reference frame is important
when testing Lorentz invariance
since the coefficients for Lorentz violation
are different in different frames.
By convention,
Lorentz tests report results
using a Sun-centered celestial equatorial frame
\cite{sunframe,km09}.
The direction of propagation $\hat v$
has Sun-frame polar angle
$\th = \text{dec.} + 90^\circ$
and Sun-frame azimuthal angle
$\ph = \text{r.a.} \pm180^\circ$,
in terms of the declination and right ascension
of the source.
The standard Sun-frame linear polarizations
$\h+$ and $\h\times$ are
related to the helicity components
through Eq.\ \rf{pols}.
The axes of the $\h+$ polarization align with
the celestial cardinal directions,
while the axes for $\h\times$ polarization
are along the intercardinal directions.

Equation \rf{hyjmf} gives the predicted
form for the  Lorentz-invariant helicity
components in the merger frame.
Before we determine the effects of Lorentz violation,
we first transform this result to the Sun frame.
This is done by passively rotating Eq.\ \rf{hyjmf} 
using the rotation operator
$R = e^{i\al_z J_z} e^{i\al_y J_y} e^{i\al_z' J_z}$,
where $\al_y$, $\al_z$, and $\al_z'$
are Euler angles relating the two frames.
The rotation acting on the spherical harmonics gives
$R\, \syjm{s}{jm}
= \sum_{m'} D^{(j)}_{m'm}(-\al_z,-\al_y,-\al_z') \syjm{s}{jm'}$,
where $D^{(j)}_{m'm}$ are the Wigner rotation matrices.
The Sun-frame Lorentz-invariant wave
can then be written as
\beq
\h{\pm2}^\text{LI}(f) 
=  u(f)\, \Y{\pm2} + u^*(-f)\, \Y{\mp2}^* \ ,
\eeq
where it is convenient to define
the direction-dependent factors
\bea
\Y{\pm2} 
&=& \sum_m  D^{(2)}_{m2}(-\al_z,-\al_y,0) \syjm{\mp2}{2m}(\th,\ph)
\notag\\
&=& \sum_m  d^{(2)}_{m2}(-\al_y) e^{i\al_zm} \syjm{\mp2}{2m}(\th,\ph) \ .
\label{Yfactors}
\eea
The $d^{(j)}_{mm'}$ in the last line are little Wigner matrices.

The $\Y{\pm2}$ factors
have helicity $\pm2$ and
account for the location ($\th$, $\ph$)
and orientation ($\al_y$,$\al_z$) of the source
in the Sun-centered frame.
While the general rotation depends on
the three Euler angles,
the $\al_z'$ angle
corresponds to a rotation about
the merger-frame $z$ axis.
This is equivalent to a change in phase
and can be absorbed into the phase constant $\ps_0$.
We may therefore take $\al_z'=0$,
leaving four angles to characterize
the location and orientation of the binary
in the Sun frame.
Note that the $\Y{\pm2}$ also
completely determine the polarization content in
the conventional limit.

Restricting attention to positive $f$
and using Eq.\ \rf{htrans},
the helicity components with Lorentz violation are
\bea
\h{\pm2} &=& u\, e^{i\de}
\big((\cos\be \mp i\cos\vt\sin\be) \Y{\pm2}
\notag \\ &&\qquad\quad
-i e^{\mp i\vp} \sin\vt \sin\be \Y{\mp2}\big) \ .
\label{h_hel}
\eea
The negative-frequency parts can be found using
the identity $\h{\pm2}(f) = \h{\mp2}^*(-f)$.
In the linear basis,
the positive-frequency parts become
\bea
\h+ &=& u\, e^{i\de}
\big((\cos\be-i\cos\vp\sin\vt\sin\be)\Y{+}
\notag \\ &&\qquad\quad
- (\cos\vt+i\sin\vp\sin\vt) \sin\be\Y{\times}\big) \ ,
\notag\\
\h\times &=& u\, e^{i\de}
\big((\cos\be+i\cos\vp\sin\vt\sin\be)\Y{\times}
\notag \\ &&\qquad\quad
+ (\cos\vt-i\sin\vp\sin\vt) \sin\be\Y{+}\big) \ ,\qquad
\label{h_lin}
\eea
where $\Y{+}$ and $\Y{\times}$
are direction factors for linear polarizations
and are related to helicity-basis factors
through $\Y{\pm2} = \Y{+}\mp i\Y{\times}$.
The linear components both obey the conjugation rule
$\h{+,\times}(f) = \h{+,\times}^*(-f)$,
which implies the time-domain $\h{+,\times}(t)$ are real,
as expected.

The signal generated in a given detector
depends on its orientation
relative to the Sun-centered frame
at the time of the observation.
The gravitational strain
for a detector can be written as
\bea
\h{\text{D}}
&=& \F+ \h+ + \F\times \h\times
\notag\\
&=& \half \F{+2}^* \h{+2} + \half \F{-2}^* \h{-2}\ .
\eea
Assuming arms of equal length,
the linear-basis antenna pattern functions are
\bea
\F+ &=& \half\big(C_{\th1}^2 - C_{\ph1}^2
- C_{\th2}^2 + C_{\ph2}^2\big)\ ,
\notag \\
\F\times &=&  C_{\th1} C_{\ph1} - C_{\th2} C_{\ph2}
\eea
where 
$C_{ak} = \hat e_a\cdot \hat l_k$
are the direction cosines
between the Sun-frame
$\hat e_\th$ and $\hat e_\ph$ vectors
and the arms of the detector,
which point along unit vectors
$\hat l_1$ and $\hat l_2$.
Assuming the $\hat l_k$ vectors are horizontal,
the cosines are
\bea
C_{\th k} &=& \cos(\ph-\al)\cos\th\cos\ch\cos\xi_k
+ \sin\th\sin\ch\cos\xi_k \
\notag \\
&& + \sin(\ph-\al)\cos\th\sin\xi_k \ ,
\notag \\
C_{\ph k} &=& \cos(\ph-\al)\sin\xi_k
- \sin(\ph-\al)\cos\ch\cos\xi_k \ , 
\eea
where $\th$ and $\ph$ are
the Sun-frame propagation angles,
$\ch$ is the colatitude of the detector,
$\al$ is the right ascension of
the laboratory zenith at the time of the detection,
and $\xi_k$ is the angle between $\hat l_k$
and local south measured to the east.
The helicity-basis pattern functions
are given by
$\F{\pm2} = \F+\mp i\F\times$.

Using either the helicity components \rf{h_hel}
or linear components \rf{h_lin},
we find that the  theoretical
positive-frequency strain with Lorentz violation
takes the simple form
\beq
\h{\text{D}} 
= A\, e^{i(\ps+\de)}
(\D^0 \cos\be - i\vec\D\cdot\hat\vs\, \sin\be) \ ,
\label{strain}
\eeq
where
\beq
\hat\vs = (\sin\vt\cos\vp,\sin\vt\sin\vp,\cos\vt)
\eeq
is the Stokes rotation axis discussed in the Appendix.
It is the normalized Stokes vector
for the faster eigenmode
and depends on the birefringence angles
$\vt$ and $\vp$ defined in Eq.\ \rf{bire_angles}.
The negative-frequency strain can be found using
$\h{D}(f) = \h{D}^*(-f)$,
ensuring that the time-domain strain is real.

We emphasize that Eq.\ \rf{strain} gives
the leading-order theoretical strain signal
for any realistic extension of linearized gravity,
including all possible Lorentz-breaking
and Lorentz-invariant modifications.

The unconventional effects in
a given event are completely determined by
the common phase $\de$, the birefringent phase $\be$,
and the $\vt$ and $\vp$ angles.
The conventional degrees of freedom
are those in the chirp amplitude $A(f)$
and phase $\ps(f)$ functions
and in the Stokes-like parameters
\bea
\D^0 &=& \half\big(\F{+2}^*\Y{+2}+\F{-2}^*\Y{-2}\big) \ ,
\notag \\
\D^1 &=& \half\big(\F{-2}^*\Y{+2}+\F{+2}^*\Y{-2}\big) \ ,
\notag \\
\D^2 &=& \tfrac{i}2\big(\F{-2}^*\Y{+2}-\F{+2}^*\Y{-2}\big) \ ,
\notag \\
\D^3 &=& \half(\F{+2}^*\Y{+2}-\F{-2}^*\Y{-2}\big) \ .
\eea
These complex factors represent a distillation
of all the usual directional degrees of freedom.

The strain
in the Lorentz-invariant case
is $\h{\text{D}}^\text{LI} = \D^0 Ae^{i\ps}$.
In this limit,
the incident polarization depends on
the location and orientation of
the source relative to the Earth
but is the same for all frequencies.
The conventional direction factor $\D^0$
accounts for the polarization
of the incident wave and
its alignment relative to the arms of the detector.

In Lorentz-violating cases with birefringence,
different frequencies can have different polarizations,
which gives additional $f$ dependence
characterized by the terms in brackets in Eq.\ \rf{strain}.
While nonbirefringent Lorentz violation
affects the phase of the signal,
birefringence can alter both
the phase and the amplitude.
The changes in amplitude can by isolated
by considering the spectral density
\begin{align}
|\h{\text{D}}|^2
&= A^2
\big(|\D^0|^2 \cos^2\be
+ |\vec\D\cdot\hat\vs|^2 \sin^2\be
\notag\\&\qquad\quad
+ \im(\D^{0*} \vec\D\cdot\hat\vs) \sin2\be)
\big)\ .
\label{spect}
\end{align}
Deviations from the expected spectral density
$|\D^0|^2A^2$ provide a generic signature
of birefringence due to Lorentz violation.

\subsection{Special cases}

Equation \rf{strain} provides
a general framework for searches
Lorentz violation in chirp signals.
While the SME describes an endless variety
of possible Lorentz violations,
experimental limitations
likely preclude a broad search.
It is therefore useful to focus on
the three main classes of violations
controlled by the three different
coefficient combinations
in Eq.\ \rf{sigma2}.
We again consider each case in turn.

\begin{figure}
  \includegraphics[width=\columnwidth]{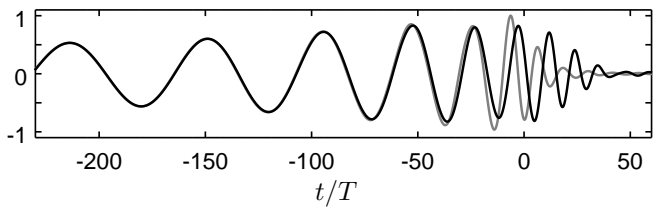}
  \vskip-10pt
  \caption{\label{disp}
    Time-domain strain for
    nonbirefringent dispersion with
    $\ta\vs^{(6)0} = 20T^3$ (black)
    and the Lorentz-invariant limit (gray).
    The Lorentz-invariant strain is generated
    using the amplitude and phase functions
    in Ref.\ \cite{ajith1}
    with mass ratio $\et=\tfrac14$.
  }
\end{figure}

{\em Nonbirefringent violations.}
Setting all but the $\kI$ coefficients
to zero yields the nonbirefringent limit.
In this limit, the strain becomes
\beq
\h{\text{D}}
= \D^0 A\, e^{i(\ps+\de)}\ .
\label{strain_nonbire}
\eeq
This simply adds the Lorentz-violating phase
$\de = (2\pi f)^{d-3}\, \ta\, \vs^{(d)0}$
to the conventional phase function $\ps$.
The modifications exist for even $d\geq 4$,
but dispersion results only when $d\geq 6$.
Figure \ref{disp} shows an example of
dispersion in the strain signal
for $d=6$ Lorentz violations.
A single point source can constrain
one direction-dependent
coefficient combination $\vs^{(d)0}$,
for fixed $d$.
Measurements from multiple sources
at different locations on the sky
could be combined to limit
the entire $\kI$ coefficient space.

A single source can, however,
constrain the isotropic limit.
At each $d$, there is one
isotropic coefficient $\kIdjm{d}{00}$.
Setting all other coefficients to zero
yields a simple one-parameter special case with
Lorentz-violating phase
\beq
\de_\text{iso} = \tfrac{1}{\sqrt{4\pi}}(2\pi f)^{d-3}\,
\ta\, \kIdjm{d}{00}\ .
\eeq
This produces polarization- and
direction-independent dispersion.
Isotropic dispersion of the form
$\om^2 = |\vec p|^2 + A |\vec p|^\al$
has been considered \cite{disp0}.
This formalism maps to SME parameters through
$\al = d-2$ and
$A = -\sqrt{1/\pi}\,\kIdjm{d}{00}$.
Note that this implies that $\al$
is an even positive integer.
Other values of $\al$ may occur
in gauge-breaking theories
or in non-field-theoretic descriptions of gravity,
but both of these possibilities likely
produce additional effects
beyond simple dispersion.
Odd $\al$ values do appear
in the CPT-odd birefringent case discussed below.
However, dispersion in that case is accompanied
by changes in polarization.

\begin{figure}
  \includegraphics[width=\columnwidth]{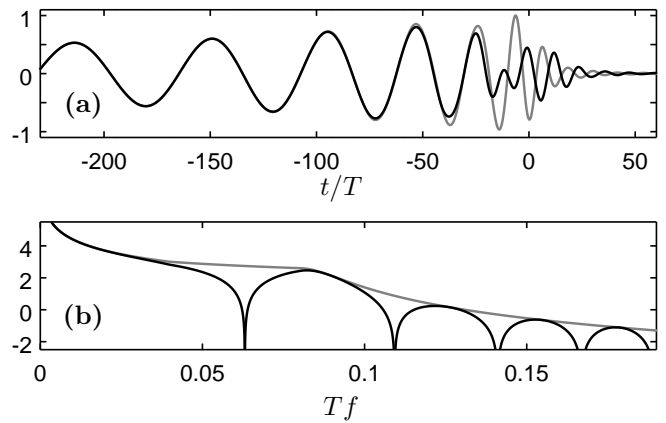}
  \vskip-5pt
  \caption{\label{d5bire}
    (a) Time-domain strain and 
    (b) base-10 logarithm of the spectral density
    for CPT-odd birefringence with
    $\ta \vs_{(0)}^{(5)} = 10T^2$ (black).
    The Lorentz-invariant limit (gray)
    is generated using
    the phase and amplitudes from
    Ref.\ \cite{ajith1}
    with mass ratio $\et=\tfrac14$.
    The merger is positioned
    so that the incident wave
    is linearly polarized in
    the Lorentz-invariant limit,
    and the detector arms are aligned
    to maximize the conventional sensitivity.
  }
\end{figure}

{\em CPT-odd birefringence.}
Taking nonzero $\kV$ coefficients
gives the CPT-odd birefringent case.
The result can be written in
term of the phase
$\de\ps = (2\pi f)^{d-3}\ta\vs_{(0)}^{(d)}$,
for odd $d\geq5$.
The positive-frequency strain reduces to
\beq
\h{\text{D}}
= A\, e^{i\ps}
(\D^0 \cos\de\ps - i\D^3 \sin\de\ps) \ ,
\label{strain_CPTodd}
\eeq
and the spectral density becomes
\begin{align}
|\h{\text{D}}|^2
&= A^2 \big(
|\D^0|^2 \cos^2\de\ps
+ |\D^3|^2 \sin^2\de\ps
\notag \\&\qquad\quad
+\im(\D^{0*}\D^3) \sin2\de\ps
\big) \ .
\end{align}
Figure \ref{d5bire} shows an example
of the theoretical strain
from $d=5$ CPT-odd birefringence.
As in the nonbirefringent case,
multiple sources at different locations
on the sky are required to fully constrain
the $\kV$ coefficient space for fixed $d$.
This case also contains an isotropic limit,
where
\beq
\de\ps_\text{iso} = \tfrac{1}{\sqrt{4\pi}}
(2\pi f)^{d-3}\ta \kVdjm{d}{00}\ .
\eeq
A single source can fully constrain this
simple special case.

{\em CPT-even birefringence.}
For the CPT-even birefringent case,
we take $\kE$ and $\kB$ coefficients to be nonzero.
For fixed $d$,
the strain is then 
given by Eq.\ \rf{strain},
with $\de=0$,
birefringent phase
\beq
\be = (2\pi f)^{d-3}\ta\big|\vs_{(+4)}^{(d)}\big| \ ,
\eeq
and
\beq
\vec\D\cdot\hat\vs = \D^1\cos\vp +\D^2\sin\vp \ .
\eeq
Using this direction factor in Eq.\ \rf{spect}
gives the spectral density.
The effects are therefore governed by
the magnitude and phase of
the coefficient combination $\vs_{(+4)}^{(d)}$.
No isotropic limit exists in this case.
This is because the $\pm4$-helicity
$\vs_{(\pm4)}^{(d)}$ combinations
couple the circular polarizations $\h{\pm2}$,
giving a change of $\pm4$ in helicity.
Only scalar helicity-zero functions
contain isotropic components,
so violations of this type are
necessarily anisotopic.

In both birefringent cases,
the unlikely possibility exists that
the observed gravitational wave is produced
in one of the propagating eigenmodes.
For the CPT-odd case,
the eigenmodes are circularly polarized,
implying that the Earth would need to lie
on the $z$ axis of the merger frame.
The wave then maintains a fixed polarization,
and the direction factors obey $\D^3=\pm\D^0$.
For the CPT-even case,
the eigenmodes are linearly polarized,
which means the Earth lies in the $x$-$y$
plane of the merger frame,
and the plane is aligned
with a polarization axis
of one of the eigenmodes.
In this scenario,
the direction factor satisfies
$\vec\D\cdot\hat\vs = \D^0$
for the fast mode
and $\vec\D\cdot\hat\vs = -\D^0$
for the slow mode.
In both the CPT-odd and CPT-even cases,
the polarization would be unaffected
and there would be no sign of Lorentz violation
in the spectral density.
However, the birefringent phase $\be$ produces dispersion
and therefore still results in a deformed strain.

\section{SUMMARY AND DISCUSSION}
\label{sec_disc}

In this paper,
we derive and study signals for Lorentz violation
in gravitational waves.
We work in the linearized-gravity limit
of the SME and consider all possible violations
that preserve the usual gauge invariance.
The violations involve $d-2$ spacetime derivatives
of the metric perturbation $h_\mn$,
where $d$ is the mass dimension
of the corresponding operator in the action,
in natural units.

Leading-order plane-wave solutions are found
in Sec.\ \ref{sec_gw}.
Lorentz violation results in two propagating modes
with conventional leading-order polarizations.
The modes can propagate at different
frequency-dependent phase velocities,
resulting in both dispersion and birefringence.
While the effects depend on the propagation direction in general,
one isotropic coefficient for Lorentz violation exists
at each dimension $d$ that gives direction-independent modifications.
Equation \rf{htrans} gives the effects
of Lorentz violation on the
circular-polarization components of a wave
after propagating an astrophysical distance
in terms of its Lorentz-invariant limit.
Equation \rf{htranslin} gives the effects
for the linear polarizations.

Signals of Lorentz violation
in waves from coalescing binaries
are derived in Section \ref{sec_bin}.
The key result is the detector-specific
theoretical strain given in Eq.\ \rf{strain}.
While the emphasis of this paper is Lorentz violation,
Eq.\ \rf{strain} describes in general
the leading-order effects
of any extension of linearized gravity,
including all possible Lorentz-violating
and Lorentz-invariant terms.

The theoretical chirp signal involves 
a number of unknown parameters that can
only be determined through observation.
The wave depends on the total binary mass $M$,
the mass ratio $\et$, and the redshift $z$.
The spins of the objects may also play a role.
There are two angles $\th$ and $\ph$
that specify the location of the binary
relative to the Earth.
Two more angles $\al_y$ and $\al_z$
characterize its orientation.

Lorentz violation introduces
a frequency-dependent common phase $\de$,
which produces dispersion but no birefringence.
It also results in a frequency-dependent
birefringent phase $\be$
that gives both dispersion and birefringence.
Birefringence produces changes in
the polarization of the wave,
which also depends on
two angles $\vt$ and $\vp$ 
that determine the polarizations
of the propagating eigenmodes.
For a given source
with propagation direction $\hat v$,
all the Lorentz-violating parameters
can be written in terms $\hat v$-dependent
combinations of spherical coefficients
for Lorentz violation using Eq.\ \rf{sigma2}.

While the formalism developed here
can be used to search
for general Lorentz violations,
forming a complete picture of a merger
is challenging even in the usual case,
due in part to the large number
of degrees of freedom.
The task can be simplified by
focusing on one of three limiting cases:
nonbirefringent violations,
CPT-odd birefringent violations,
and CPT-even birefringent violations.

Nonbirefringent dispersion
results from the $\kI$ coefficients
for Lorentz violation.
Dispersion deforms chirp waveforms
by adding unconventional frequency dependence
$\sim f^{d-3}$
to the phase.
A single source can be used to limit
the direction-dependent coefficient combination
$\vs^{(d)0}$ in Eq.\ \rf{sigma2}
for $d=6,8,\ldots$.
Results from multiple sources at different locations
on the sky could be combined to constrain
the $\kI$ spherical coefficients
for Lorentz violation for fixed dimension $d$.

While birefringent Lorentz violation comes
with dispersion,
a key signature differentiating
it from the nonbirefringent case
is a change a polarization that evolves
as the wave propagates.
The conventional case is expected
to give a frequency-independent polarization,
so a polarization that changes with frequency
is an indicator for birefringent Lorentz violation.

While a single detector measures a single polarization component,
a wave's polarization can in principle be reconstructed
by combining the strain data from multiple detectors.
Alternatively, signs of birefringence
can be found in the spectral density.
The spectral density of the strain
is insensitive to the phase of incident wave
and is therefore insensitive to dispersion.
Changes in polarization due to birefringence 
distort a detectors response to a wave,
affecting the strain amplitude.
As a result, the spectral density is highly
sensitive to frequency dependence in the polarization.
The general predicted spectral density
with Lorentz violation is given in Eq.\ \rf{spect}.
A deviation for the expected power law
during the inspiral could be a signal of birefringence,
for example.

CPT-odd and CPT-even birefringence each
produce distinctive changes in the polarization.
The $\kV$ coefficients give CPT-odd birefringence,
which leads to  a simple frequency-dependent
rotation of the polarization
about the propagation direction.
The result is a change in the polarization angle
but no change in the degree of linear or circular polarization.
CPT-even birefringence stems from the $\kE$ and $\kB$
coefficients for Lorentz violation.
The effects in the CPT-even case are more complicated,
giving a change in the polarization angle and
the degrees of linear and circular polarization.

Because the effects of Lorentz violation
depend on the direction of propagation
and the polarization,
some waves may experience minimal defects
even if Lorentz violation is significant in general.
This leads to a potential selection bias 
that is common in other astrophysical tests.
Dispersion and birefringence may distort
a gravitational wave to the point where
it is no longer recognizable as a potential chirp event.
The detected waves may be those
that happen to propagate in particular directions
and have particular polarizations
that produce very little change.
This possibility can be ruled out experimentally
by constraining the underlying coefficients
for Lorentz violation
through observations of multiple
sources at different points on the sky
and with different polarizations.

The precision at which Lorentz violation
can be tested in a gravitational wave
is largely determined by the effective
propagation time $\ta$
and the chirp time constant $T$.
The effects of Lorentz violation may
be significant when the 
phase in Eq.\ \rf{phase}
is of order one.
Frequencies up to $\sim 1/T$ contribute
to the chirp,
so gravitational waves are expected
to test combinations of coefficients
for Lorentz violation
at levels approaching $\sim T^{d-3}/\ta$.
Assuming propagation distances
on the order of a Gpc
and a total mass of around fifty
solar masses,
this gives an approximate
sensitivity at the level of
$\sim 10^{-16}$ m to $d=5$ coefficients
and
$\sim 10^{-11}$ m$^2$ to $d=6$ coefficients.

At present, few gravitational-wave bounds exist
on the spherical coefficients for Lorentz violation.
Estimated limits on birefringence in GW150914
have been used to bound
one combination of $d=5$ CPT-odd $\kVdjm{5}{jm}$ coefficients
at the level of $10^{-14}$ m
and one combination of $d=6$ CPT-even
$\kEdjm{6}{jm}$ and $\kBdjm{6}{jm}$ coefficients
at the level of $10^{-8}$ m$^2$
\cite{kmgw1}.
More sophisticated analyses will likely
achieve sensitivities 
orders of magnitude
beyond these bounds.

The best constraints on $\kIdjm{6}{jm}$
coefficients come from the absence of
gravitational \v Cerenkov radiation
in high-energy cosmic rays \cite{cerenkov2}.
Lorentz violation in gravity 
can lead to subluminal wave speeds.
High-energy particles traveling
faster than gravity will radiate
gravitational waves, losing energy.
Observations of high-energy cosmic
rays place stringent limits on
the coefficients for Lorentz violation
responsible for the changes in velocity.
The high energies involved mean
Lorentz invariance is tested
at much higher frequencies,
giving sensitivities that are many orders
of magnitude beyond what can be achieved
through observations of low-frequency mergers.
However, \v Cerenkov studies generally
make a number of simplifying assumptions
concerning Lorentz violation in
the cosmic-ray particles
and their interactions with gravity,
assumptions that may not hold in nature.
By contrast,
dispersion and birefringence provide 
clean signatures of Lorentz violation in pure gravity,
completely independent of violations in other sectors.

Tests of short-range gravity have also placed
constraints on $d=6$ Lorentz violations
\cite{shortrange1}.
Binary-merger observations are expected to give
sensitivities to $d=6$ violations that are
several orders of magnitude better than short-range tests.
Due to the long wavelengths of the gravitational waves
and the sub-millimeter reach of the laboratory tests,
short-range gravity experiments will likely achieve
better sensitivities to higher-$d$ violations.
Note, however, that these tests should be viewed as complementary
since gravitational waves provide access to
the ``vacuum'' coefficients in Table \ref{kcoeffs},
while short-range experiments are sensitive
to a different set of ``Newton'' coefficients $\kN$.
Each of these sets of coefficients
are different combinations
of the underlying coefficients
for Lorentz violation in Eq.\ \rf{sqk},
so the two classes of experiment test fundamentally
different forms of Lorentz violation.

\section*{Acknowledgments}

This work was supported in part 
by the United States National Science Foundation 
under grant number PHY-1819412.

\section*{APPENDIX: STOKES PARAMETERS}
\label{sec_stokes}

As with electromagnetic radiation,
one can define gravitational Stokes parameters
that characterize the power associated
with different polarizations.
They also provide a simple picture
for the effects of birefringence.
While birefringence generally produces
complicated changes in the polarization of a wave,
the effects can be understood as a simple rotation
of the Stokes parameters \cite{km09}. 

While gauge-invariant extensions of general relativity
can in principle
give up to 6 polarizations, resulting in
36 gravitational Stokes parameters \cite{stokes},
the limit considered in this work
produces two propagating modes
with conventional leading-order polarizations.
For positive frequencies,
we define four real gravitational Stokes parameters 
\bea
S^0 &=& |\h+|^2  + |\h\times|^2
= \half\big(|\h{+2}|^2 + |\h{-2}|^2\big) \ ,
\notag \\
S^1 &=& |\h+|^2  - |\h\times|^2
= \re\big(\h{+2}^*\h{-2}\big) \ ,
\notag \\
S^2 &=& 2\re\big(\h+^*\h\times\big)
= \im\big(\h{+2}^*\h{-2}\big) \ ,
\notag \\
S^3 &=& 2\im\big(\h+^*\h\times\big)
= \half\big(|\h{+2}|^2 - |\h{-2}|^2\big) \ .
\eea
The Stokes parameters obey $(S^0)^2 = \vec S^{\,2}$,
where $\vec S = (S^1,S^2,S^3)$
is the Stokes vector.
The Stokes vector can also be written
in terms of $\pm4$-helicity components
$S_{(\pm 4)} = S^1\mp iS^2$
and a 0-helicity component
$S_{(0)} = S^3$.

Some understanding of the Stokes parameters
can be gained by parameterizing
an arbitrary polarization
using the form
\beq
\h{\pm2} = a
\big(\cos\tfrac{\ch}{2}\pm
\sin\tfrac{\ch}{2}\big) e^{\mp i\ze/2} \ ,
\eeq
where $-\tfrac\pi2\leq\ch\leq\tfrac\pi2$
and $0\leq\ze< 2\pi$.
In the linear basis,
we can write this as
\beq
\begin{pmatrix}
  \h+\\\h\times
\end{pmatrix}
= a
\begin{pmatrix}
  \cos\tfrac\ze2& -\sin\tfrac\ze2\\
  \sin\tfrac\ze2& \cos\tfrac\ze2
\end{pmatrix}
\begin{pmatrix}
  \cos\tfrac\ch2\\
  i\sin\tfrac\ch2
\end{pmatrix} \ .
\eeq
This gives general elliptical
polarization along rotated axes
$\hat e_1 = \cos\tfrac\ze4\hat e_\th + \sin\tfrac\ze4\hat e_\ph$
and
$\hat e_2 = \cos\tfrac\ze4\hat e_\ph - \sin\tfrac\ze4\hat e_\th$,
implying that $\tfrac\ze4$ is the linear polarization angle.
The angle $\ch$ determines the degree of circular polarization,
with $\ch=0$ for linear,
$\ch=\tfrac\pi2$ for right-handed circular,
and
$\ch=-\tfrac\pi2$ for left-handed circular
polarizations.

In terms of $\ze$ and $\ch$,
the Stokes parameters are
\beq
S^0 = |a|^2\ ,\ \
\vec S = S^0
\big(\cos\ch\cos\ze,\cos\ch\sin\ze,\sin\ch\big) \ .
\eeq
Each Stokes vector $\vec S$ defines
a unique point on a sphere of radius $S^0$
analogous to the Poincar\'e sphere from optics.
Every point on the sphere represents
a unique polarization.
Points in the $S^1$-$S^2$ plane are
the linear polarizations,
with $\h+$ on the positive $S^1$ axis
and $\h\times$ on the negative $S^1$ axis.
The upper hemisphere, with $S^3>0$,
contains all right-handed elliptical polarizations,
and the lower hemisphere gives
left-handed elliptical polarizations.
The poles correspond to the two circular polarizations.
In general, orthogonal polarizations point to
opposite points on the sphere.
The degree of linear polarization is $\cos\ch$,
and $\sin\ch$ gives the degree of circular polarization.

Assuming Eq.\ \rf{hyjmf},
the Lorentz-invariant Stokes parameters
for a merger are
\bea
S^0_\text{LI} &=& \half |u|^2 (\cos^8\tfrac{\th}{2} + \sin^8\tfrac{\th}{2}) \ ,
\notag \\
S^1_\text{LI} &=& |u|^2 \sin^4\tfrac{\th}{2} \cos^4\tfrac{\th}{2}\ ,
\notag \\
S^2_\text{LI} &=& 0 \ ,
\notag \\
S^3_\text{LI} &=& \half |u|^2 (\cos^8\tfrac{\th}{2} - \sin^8\tfrac{\th}{2}) \ ,
\eea
in the merger frame.
Note that this only depends on the merger-frame polar angle $\th$.
This shows that
we get right-handed circular polarization
along the $+z$ merger axis
and left-handed circular polarization
along the -$z$ axis.
Waves traveling in the $x$-$y$ plane
are linearly polarized.
All linear and elliptical polarizations
have a polarization angle of $\tfrac\ze4=0$.
In the Sun frame,
the Stokes parameters are
$S^\mu_\text{LI} = |u|^2 S^\mu_{\mathcal Y}$,
where $S^\mu_{\mathcal Y}$
are the Stokes parameters constructed from
the direction factors in Eq.\ \rf{Yfactors}.

The Lorentz-violating parameters
in Eq.\ \rf{sigma1} can be thought of
as conveniently normalized
Stokes parameters for the birefringent eigenmodes.
The vector
\beq
\vec\vs = \big(
  \half(\vs_{(+4)}+\vs_{(-4)}),\
  \tfrac{i}{2}(\vs_{(+4)}-\vs_{(-4)}),\
  \vs_{(0)}\big)
\label{sigmavector}
\eeq
points in the direction of the Stokes vector
for the faster eigenmode
and opposite the vector for the slower mode.
Note, however, that $(\vs^0)^2 \neq \vec\vs^{\,2}$.

Changes in polarization due to birefringence
will cause the Stokes vector $\vec S$ of a wave
to evolve as it propagates.
Only waves with $\vec S$ along $\vec\vs$,
corresponding to one of the eigenmodes,
will remain unaltered.
To find the effects for other polarizations,
we define the orthonormal Stokes basis
\bea
\hat\vs &=&
\vec\vs/|\vec\vs|
=(\sin\vt\cos\vp,\sin\vt\sin\vp,\cos\vt)\ ,
\notag \\
\hat\vt &=&
(\cos\vt\cos\vp, \cos\vt\sin\vp, -\sin\vt)\ ,
\notag \\
\hat\vp &=&
(-\sin\vp,\cos\vp, 0)\ ,
\label{sigmahat}
\eea
in terms of the birefringence angles $\vt$ and $\vp$.
Expanding an arbitrary Stokes vectors in this basis,
$\vec S = S^\vs \hat\vs + S^\vt \hat\vt + S^\vp \hat\vp$,
we arrive at a set of Stokes parameters 
associated with the eigenmodes,
\begin{gather}
S^\vs = \half\big(|\h{f}|^2 - |\h{s}|^2\big)\ ,
\notag \\
S^\vt = \re\big(\h{f}^*\h{s}\big)\ , \quad
S^\vp = \im\big(\h{f}^*\h{s}\big)\ .
\end{gather}
An eigenmode differs from its Lorentz-invariant limit
by a phase,
$\h{f,s} = e^{i\de \mp i\be} \h{f,s}^\text{LI}$,
giving
\bea
S^\vs &=& S^\vs_\text{LI} \ ,
\notag\\
S^\vt &=& \cos(2\be) S^\vt_\text{LI}
-\sin(2\be) S^\vp_\text{LI} \ ,
\notag\\
S^\vp &=& \cos(2\be) S^\vp_\text{LI}
+\sin(2\be) S^\vt_\text{LI} \ .
\eea
This shows that the Stokes vector
with Lorentz violation $\vec S$
can be obtained from the Stokes vector
without Lorentz violation $\vec S_\text{LI}$
by rotating $\vec S_\text{LI}$
about the axis $\hat\vs$ by angle $2\be$,
which is the change in relative phase due to birefringence.

In the CPT-odd case,
where $\vt = 0$ or $\vt=\pi$,
the rotation axis $\hat\vs$ points
to one of the poles of the Poincar\'e sphere.
Birefringence causes a rotation about the $S^3$ axis
by $\de\ze = \pm 2\be$,
changing the linear polarization angle
by $\pm\be/2$.

For the CPT-even birefringence,
where $\vt = \pi/2$,
the rotation axis $\hat\vs$ lies in
the $S^1$-$S^2$ plane
at an angle $\vp$ from the $S^1$ axis.
Unless $\vec S_\text{LI}$ happens to align with $\hat\vs$,
the Stokes vector rotates
on a cone centered around $\hat\vs$.
This causes changes in both the $\ze$ and $\ch$
polarization angles.
The linear polarization angle changes,
as do the degree of linear polarization
and degree of circular polarization.

The Stokes parameters can also be used
to track the evolution of the polarization
as it propagates from the source to the observer.
The infinitesimal change in phase
$d\ps_\pm = d\ps_0 + \om (-\vs^0 \pm |\vec\vs\,|)dl$
changes the eigenmodes by
$d\h{f,s} = -id\ps_\pm \h{f,s}$.
The result is an infinitesimal rotation
of the Stokes parameters about $\hat\vs$.
The rotation can be written
\beq
\frac{d\vec S}{dl} = 2 \om \vec\vs\times\vec S \ ,
\eeq
giving right-handed rotations of $\vec S$
about $\hat\vs$ at a rate of $2\om|\vec\vs|$.

\end{document}